# Formation of a Te-Ag Honeycomb Alloy: A New Type of Two-Dimensional Material

J. Shah, H.M. Sohail, R.I.G. Uhrberg, and W. Wang*

*Department of Physics, Chemistry, and Biology, Linköping University, S-581 83 Linköping, Sweden*

**Abstract**

Inspired by the unique properties of graphene, the focus in the literature is now on investigations of various two-dimensional (2D) materials with the aim to explore their properties for future applications. The group IV analogues of graphene, i.e., silicene, germanene and stanene have been intensively studied in recent years. However, their semi-metallic band structures hamper their use in electronic applications. Hence, the synthesis of 2D materials with band gaps of various sizes has attracted a large interest. Here, we report a successful preparation of a 2D Te-Ag binary alloy with a honeycomb structure. Angle-resolved photoelectron spectroscopy (ARPES) in combination with first-principles calculations using density functional theory (DFT) confirmed the formation of this binary alloy. The semiconducting property is verified by the ARPES data and a direct gap of ~0.7 eV is predicted by the DFT calculations.

**Introduction**

The research field of two-dimensional (2D) materials has attracted great attention in recent years. Such materials may have unique properties that are not present in the three-dimensional (3D) bulk. The search for 2D materials for applications in the next generation of electronic and optoelectronic devices is currently quite intense. The most well known example is the discovery of graphene[1] and subsequent investigations of single element graphene-analogues from group IV, such as silicene, germanene, and stanene[2]. Another type of 2D material consists of high-Z (Z=atomic number) surface alloys on noble metals. Surface alloying is recognized as a viable way to achieve unique physical and chemical properties not found in the bulk. For example, a giant Rashba-type of spin splitting[3] was observed for $Ag_2Bi$[4], while the corresponding replacement of Ag by the lighter element Sb resulted in a barely observable split[5]. Very recently, a 2D Sb analogue of graphene, so called antimonene, was reported to be successfully formed on Ag(111), with a lattice constant and orientation matching a ($\sqrt{3}\times\sqrt{3}$)R30° supercell[6]. Hence, it is very interesting and worthwhile to examine whether Te, the group VI neighbor of Sb, forms a graphene-like flat 2D honeycomb structure or a surface alloy with spin splitting on Ag(111). Actually, in this paper, we report experimental and theoretical results that provide evidence for the combination of the two alternatives. We find that the electronic band structure matches that of a binary Te-Ag 2D honeycomb alloy, which is a new type of structure in the family of 2D materials

From the literature, Te and Ag are known to form an alloy usually referred to as silver telluride ($Ag_2Te$), which is a narrow gap (40 – 170 meV) semiconductor in the bulk form with a large magnetoresistance after doping[7,8]. Vacuum-evaporated thin films and nanowires of $Ag_2Te$ have attracted attention because of their high Seebeck coefficient of the thermoelectric effect[9]. However, to the best of our knowledge, a single Te-Ag layer or an $Ag_2Te$ surface alloy has not been discussed yet in the literature.

**Sample preparation**

In order to prepare the one atomic layer thick Te-Ag alloy, presented in this paper, a Ag(111) crystal was cleaned by repeated cycles of sputtering by Ar+ ions (1 keV) and annealing at approximately 400 °C until a sharp (1×1) low energy electron diffraction (LEED) pattern was obtained. Tellurium was deposited onto the Ag(111) sample at room temperature from a source that was calibrated by a thickness monitor. The data were obtained from a sample prepared by depositing 1/3 monolayer (ML) of Te followed by annealing at approximately 200 ˚C for 5 minutes. After the preparation, the sample showed a sharp √3×√3 LEED pattern with respect to Ag(111), see Fig. 1(a), which indicates an ordered structure with a well-defined periodicity on the surface. Core-level spectroscopy provides information about the inner shell electrons, which in turn can be used to gain information about the atomic structure. Fig. 1(b) shows a spectrum of the Te 4d core-level. The spectrum can be well fitted by just one 4d spin-orbit split component plus a linear background. This indicates that most Te atoms are located at identical positions.

**Results and discussion**

The band structure of the Te-Ag sample, probed by angle resolved photoelectron spectroscopy (ARPES), is presented in Figs. 2(a) and 2(b). ARPES data reveal a semiconducting band structure in agreement with the symmetric line shape of the Te 4d spectrum. The band dispersions follow a √3×√3 periodicity, which is clear from the band mapping along the $\overline{\Gamma}\overline{K}\overline{M}\overline{\Gamma}$ and $\overline{\Gamma}\overline{M}\overline{\Gamma}$ paths of the √3×√3 surface Brillouin zone (SBZ), indicated in Fig. 2(c). From these data, one can conclude that there are two bands separated in energy and momentum space. The upper band, S1, has a maximum at a binding energy ($E_B$), with respect to the Fermi level ($E_F$), of ~0.75 eV at $\overline{\Gamma}_1$. The dispersion of S1 shows a local maximum as well as a local minimum at $\overline{M}$ along $\overline{K}\overline{M}\overline{K}$ and $\overline{\Gamma}\overline{M}\overline{\Gamma}$, respectively. This makes $\overline{M}$ a saddle point with a diverging density of states (van Hove singularity). The dispersion of S2 shows an inverted parabolic shape with the maximum at an $E_B$ of 1.1 eV at $\overline{\Gamma}_1$. In contrast to S1, the S2 band exhibits a small splitting along both $\overline{\Gamma}\overline{M}$ and $\overline{\Gamma}\overline{K}$, indicated by the black arrows. As discussed in the literature, high-Z surface alloys on noble metal surfaces show a Rashba-type of spin splitting, which looks quite similar to the split of S2. Moreover, the smallest splitting found on an Ag(111) surface alloy is for $Ag_2Sb$. Since the atomic number of Te is increased one step compared to Sb, one can expect to observe a spin splitting.

Based on the number of bands, the (√3×√3)R30˚ periodicity, and the line shape of the Te 4d spectrum, some more or less likely models can be envisioned. We have considered four models with a √3×√3 supercell with respect to Ag(111). i) Te adatoms (1/3 ML) positioned at hollow sites on the Ag(111) surface. ii) A monolayer of graphene-like Te, i.e., tellurene. iii) A substitutional $Ag_2Te$ surface alloy, i.e., one out of three Ag atoms in the √3×√3 cell is substituted by Te in the upper atomic layer. iv) A Te-Ag 2D binary alloy forming a honeycomb structure. These four models are visualized after full relaxation in Fig. S1 of the supplementary information. We find that only the last model can explain all the features observed in the ARPES data. The theoretical band structures of all four models are presented together with the experimental one in Fig. S2.

In the following we present band structure calculations, which were performed based on DFT using the PAW method[10] implemented in VASP[11]. We employed the generalized gradient approximation (GGA) with the PBE[12] exchange-correlation. Our calculated results lead to the conclusion that the Te-Ag alloy is well described by model iv) mentioned above and shown in Fig. 3(a). One Te atom (blue ball) and one Ag atom (yellow ball) form a honeycomb structure with a lattice constant of 5.0 Å, which matches the √3×√3 supercell of the Ag(111) surface. After full relaxation of the slab described in the methods part, the Te-Ag binary surface alloy was stable on the Ag(111) surface with very small buckling (0.1 Å). Figures 3(b) and 3(c) show the results of band structure calculations, plotted along $\overline{M}\overline{\Gamma}\overline{K}\overline{M}$, without and with spin orbit coupling, respectively. Red dots highlight p states of the Te-Ag binary alloy. Comparing Figs. 2 and 3, one finds a striking agreement between the ARPES data and the calculated band structure with spin orbit coupling. The calculated $\Sigma_1$ and $\Sigma_2$ bands in Fig. 3(c) reproduce the dispersions of S1 and S2 extremely well. There is an energy difference of ≈0.31 eV between $\Sigma_1$ and $\Sigma_2$ at the $\overline{\Gamma}$ point which closely reproduces the experimental separation of ≈0.36 eV between S1 and S2. Regarding the absolute energy positions of $\Sigma_1$ and $\Sigma_2$ compared to S1 and S2 we find a difference of ~0.3 eV. A most likely reason for the deviation is that self-energy effects, to obtain the quasiparticle band structure, are not included in the present calculations[13]. Another extraordinary agreement is the small but resolvable splitting of $\Sigma_2$ appearing after applying spin orbit interaction in the calculation. The combination of spin-orbit coupling and the broken inversion symmetry at the surface point to a split of the Rashba type of $\Sigma_2$/S2. It is interesting to note that the calculated value of the direct band gap at $\bar{\Gamma}$ in Fig. 3(c) is ~0.65 eV, which is very close to the ~0.75 eV binding energy of S1 in Fig. 2. This strongly indicates that the Fermi level of the system is actually pinned at a position very close to the conduction band minimum of the Te-Ag alloy.

**Conclusions**

We have successfully grown a 2D binary Te-Ag alloy with a honeycomb structure on Ag(111). The atomic and electronic structures were experimentally examined by LEED, core-level spectroscopy and ARPES. The proposed atomic model was verified by comparing the theoretical band structure from first principles DFT calculations with the experimental one obtained by ARPES. A small Rashba-type of splitting was observed for the lower S2 band. Our results on this new type of 2D material should inspire further investigations on honeycomb structured binary alloys.

**Methods**

**Experiments**. Samples were prepared in-situ in an ultrahigh vacuum (UHV) system equipped with LEED and ARPES. ARPES and core-level data were obtained at beam line I4 at the MAX-lab synchrotron radiation facility. The data were acquired at room temperature by a Phoibos 100 analyzer from Specs with a two-dimensional detector. The energy and angular resolutions were 50 meV and 0.3˚, respectively.

**Calculations**. First-principles density functional theory (DFT) calculations were used to interpret the experimental electronic structure data from ARPES. Atomic structures were modeled by a slab, which was built from nine Ag layers terminated by a Te containing √3×√3 supercell. Four different structures were considered, see Figure S1 in supplementary material. About 19 Å of vacuum spacing was used to avoid interaction between neighboring slabs of the periodic structure. The positions of all atoms were fully relaxed using the functional of Perdew, Burke and Ernzerhof (PBE) and the projector augmented wave (PAW) method including van der Waals (vdW) interaction within Vienna ab initio simulation package (VASP) code. The energy cutoff of the plane-wave basis set was 375 eV, and the k-point mesh was (9×9×1). All atoms were relaxed until the average force was within 0.01 eV/Å. Band structure calculations were implemented considering spin-orbital coupling.

**Acknowledgements**

Technical support from Dr. Johan Adell, Dr. Craig Polley and Dr. T. Balasubramanian at MAX-lab is gratefully acknowledged. Financial support was provided by the Swedish Research Council (Contract No. 621-2014-4764) and by the Linköping Linnaeus Initiative for Novel Functional Materials supported by the Swedish Research Council (Contract No. 2008-6582).

**Author contributions**

J.S. conducted the experiments together with H.M.S., and W.W., R.I.G.U. did the theoretical calculations and wrote the paper. All authors took part in data analysis. R.I.G.U. is the group leader.

**Supplementary Information**

Atomic structures and calculated band structures of models i) - iv).

**Competing interests**

The authors declare no competing interests.

**Data availability statement**

The datasets generated during and/or analysed during the current study are available from the corresponding author on reasonable request.

**References**

1. Geim, A. K. & Novoselov, K. S. The Rise of Graphene. *Nature Mater.* **6**, 183 (2007)

2. Krawiec, M. Fuctionalization of Group-14 Two-Dimensional Materials. *J. Phys.: Condens. Matter.* **30**, 233003 (2018), and references therein.

3. Bychkov, Y.A. & Rashba, E.I. Properties of a 2D Electron Gas with Lifted Spectral Degeneracy. *JETP Lett*. **39**, 78 (1984)

4. Ast, C. R., Henk, J., Ernst, A., Moreschini, L., Falub, M. C., Pacilé, D., Bruno, P., Kern, K., Grioni, M. Giant Spin Splitting through Surface Alloying. *Phys. Rev. Lett*. **98**, 186807 (2007)

5. Moreschini, L., Bendounan, A., Gierz, I., Ast, C. R., Mirhosseini, H., Höchst, H., Kern, K., Henk, J., Ernst A., Ostanin, S., Reinert, F., Grionial, M. Assessing the Atomic Contribution to the Rashba Spin-Orbit Splitting in Surface Alloys: Sb/Ag(111). *Phys. Rev. B* **79**, 075424 (2009)

6. Shao, Y., Liu, Z.-L., Cheng, C., Wu, X., Liu, H., Liu, C., Wang, J.-O., Zhu, S.-Y., Wang, Y.-Q., Shi, D.-X., Ibrahim, K., Sun, J.-T., Wang, Y.-L., Gao, H.-J. Epitaxial Growth of Flat Antimonene Monolayer: A New Honeycomb Analogue of Graphene. *Nano Lett*. **18**, 2133 (2018)

7. Xu, R., Husmann, A., Rosenbaum, T. F., Saboungi, M. L., Enderby, J. E., Littlewood, P. B. Large Magnetoresistance in Non-Magnetic Silver Chalcogenides. *Nature* **390**, 57 (1997)

8. Chuprakov, I. S., Dahmen, K. H. Large Positive Magnetoresistance in Thin Films of Silver Telluride. *Appl. Phys. Lett.* **72**, 2165 (1998)

9. Li, F., Hu, C., Xiong, Y., Wan, B., Yan, W., Zhang, M. Phase-Transition-Dependent Conductivity and Thermoelectric Property of Silver Telluride Nanowires. *J. Phys. Chem. C* **112**, 16130 (2008)

10. Kresse, G., Joubert, D. From Ultrasoft Pseudopotentials to the Projector Augmented-Wave Method, *Phys. Rev. B* **59**, 1758 (1999)

11. Perdew, J. P., Burke, K., Ernzerhof, M. Generalized Gradient Approximation Made Simple, *Phys. Rev. Lett.* **77**, 3865 (1996)

12. Kresse, G., Furthmüller, J. Efficient Iterative Schemes for *Ab Initio* Total-Energy Calculations using a Plane-Wave Basis Set, *Phys. Rev. B* **54**, 11169 (1996)

13. Aryasetiawan, F., Gunnarsson, O. The GW Method. *Rep. Prog. Phys.* **61**, 237 (1998)

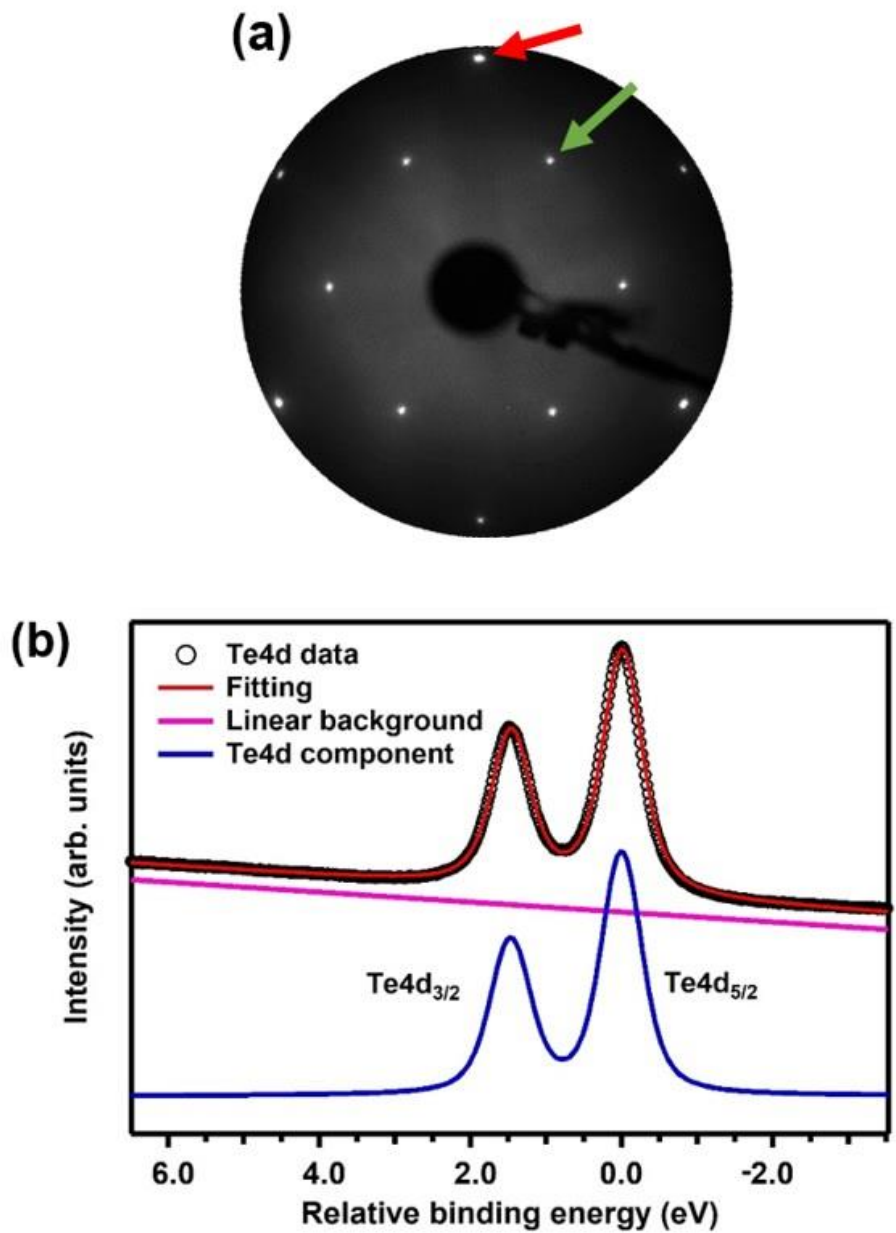

**Figure 1**. **LEED and core-level spectroscopy data.** (a) LEED pattern (78 eV) of the Te-Ag sample showing sharp diffraction spots, indicating a well-ordered structure with a (√3×√3)R30° periodicity relative to Ag(111). One of the Ag 1×1 spots is indicated by a red arrow. The green arrow points out one of the √3×√3 spots. (b) Te 4d core-level spectrum obtained using a photon energy of 80 eV at normal emission. The black circles are the experimental data and the fitting curve is the sum of one 4d spin-orbit split component and a linear background. Fitting parameters: Spin-orbit split: 1.47 eV, Branching ratio: 0.636, Gaussian width: 387 meV, and Lorentzian width: 373 meV. The asymmetry parameter of the Doniach–Šunjić line profile is 0, which indicates a semiconducting character.

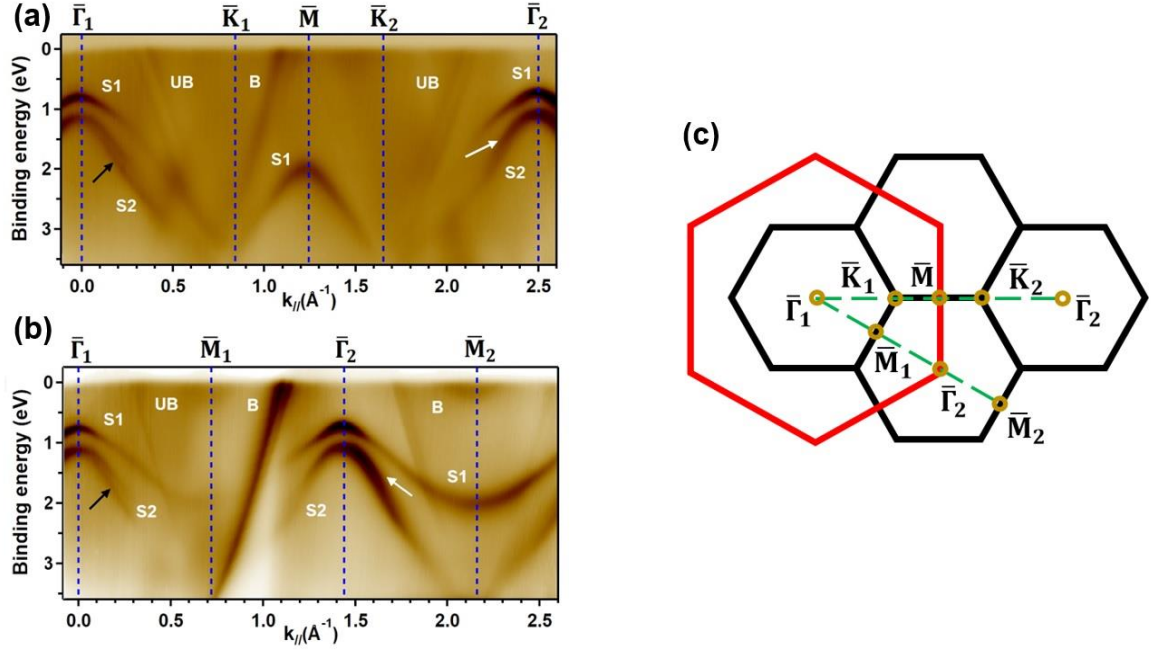

**Figure 2. Electronic band structure of the Te-Ag alloy on Ag(111) obtained by ARPES.** (a) Energy bands mapped along the $\overline{\Gamma}\overline{K}\overline{M}\overline{K}\overline{\Gamma}$ line of the SBZ. (b) Energy bands mapped along $\overline{\Gamma}\overline{M}\overline{\Gamma}\overline{M}$. The photon energy was 35 eV. Two bands are indicated by S1 and S2, respectively, in (a) and (b). These bands constitute the 2D band structure of the Te-Ag alloy along the high symmetry lines. B originates from direct transitions involving Ag sp bulk bands. UB denotes umklapp scattering of B by √3×√3 reciprocal lattice vectors. Black arrows indicate band splittings, interpreted to be of the Rashba type. White arrows point to places in the band structure, where the splitting is expected, but is less clear in the data. (c) Schematic drawing of SBZs. Red hexagon shows the 1×1 SBZ of the Ag(111) surface. Black hexagons represent √3×√3 SBZs of the Te-Ag alloy. Green dashed lines indicate the high symmetry lines in the SBZs, along which the experimental data were obtained. Several symmetry points are marked by yellow circles.

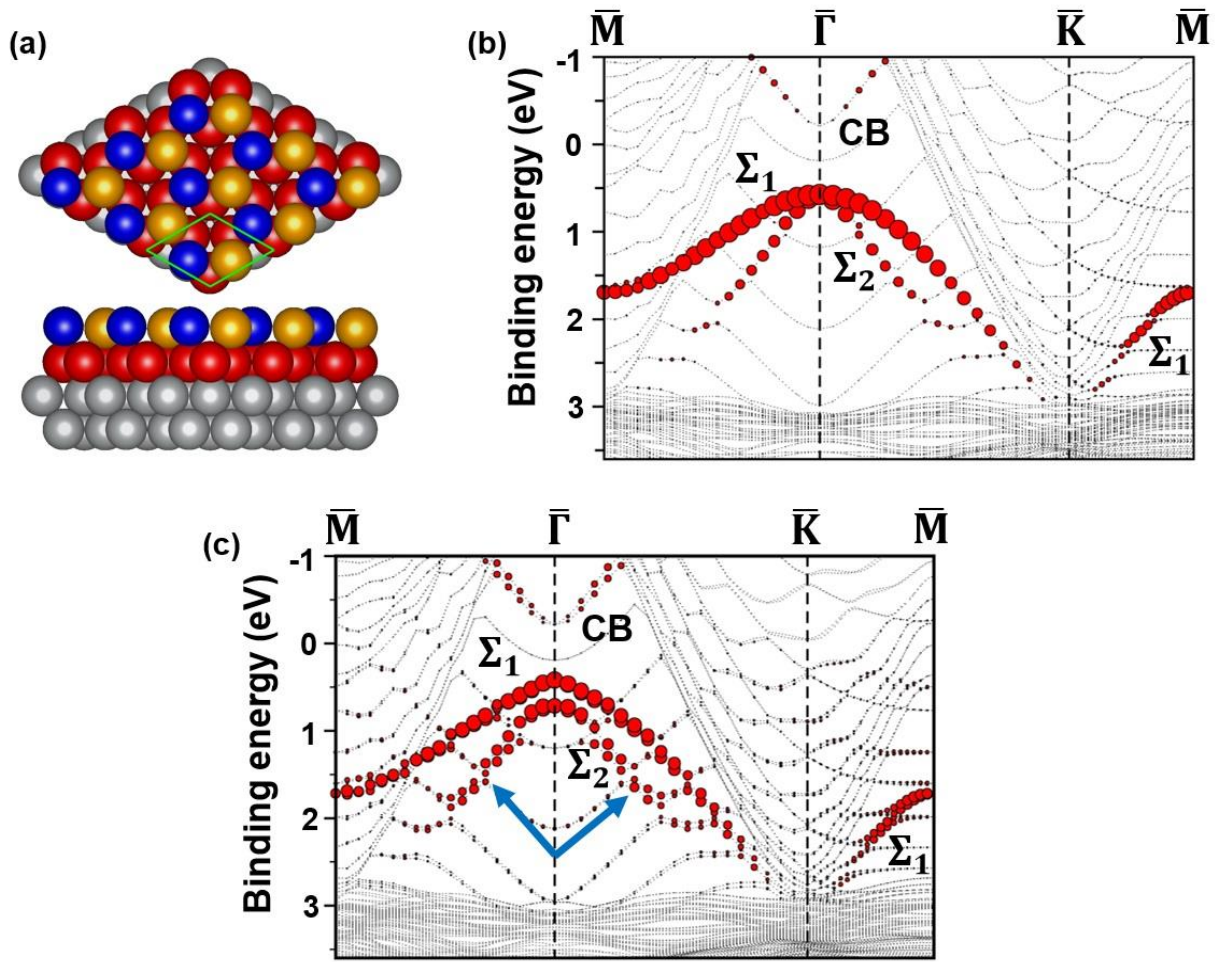

**Figure 3. Atomic model and band structure of the Te-Ag binary alloy on Ag(111) calculated using PBE+DFT.** (a) Top and side views of the atomic model of the Te-Ag binary alloy on Ag(111). The blue balls represent Te atoms. Yellow balls represent Ag atoms which form an alloy with Te atoms. Red balls represent the first layer of the Ag(111) slab. Grey balls represent deeper Ag layers. (b) and (c) The calculated band structure plotted along $\overline{M}\overline{\Gamma}\overline{K}\overline{M}$, without and with spin orbit coupling, respectively. Two valence bands from the 2D Te-Ag alloy are labeled $\Sigma_1$ and $\Sigma_2$ and the conduction band is labeled CB. The blue arrows point at regions where the $\Sigma_2$ band shows a small split.

# Supplementary information

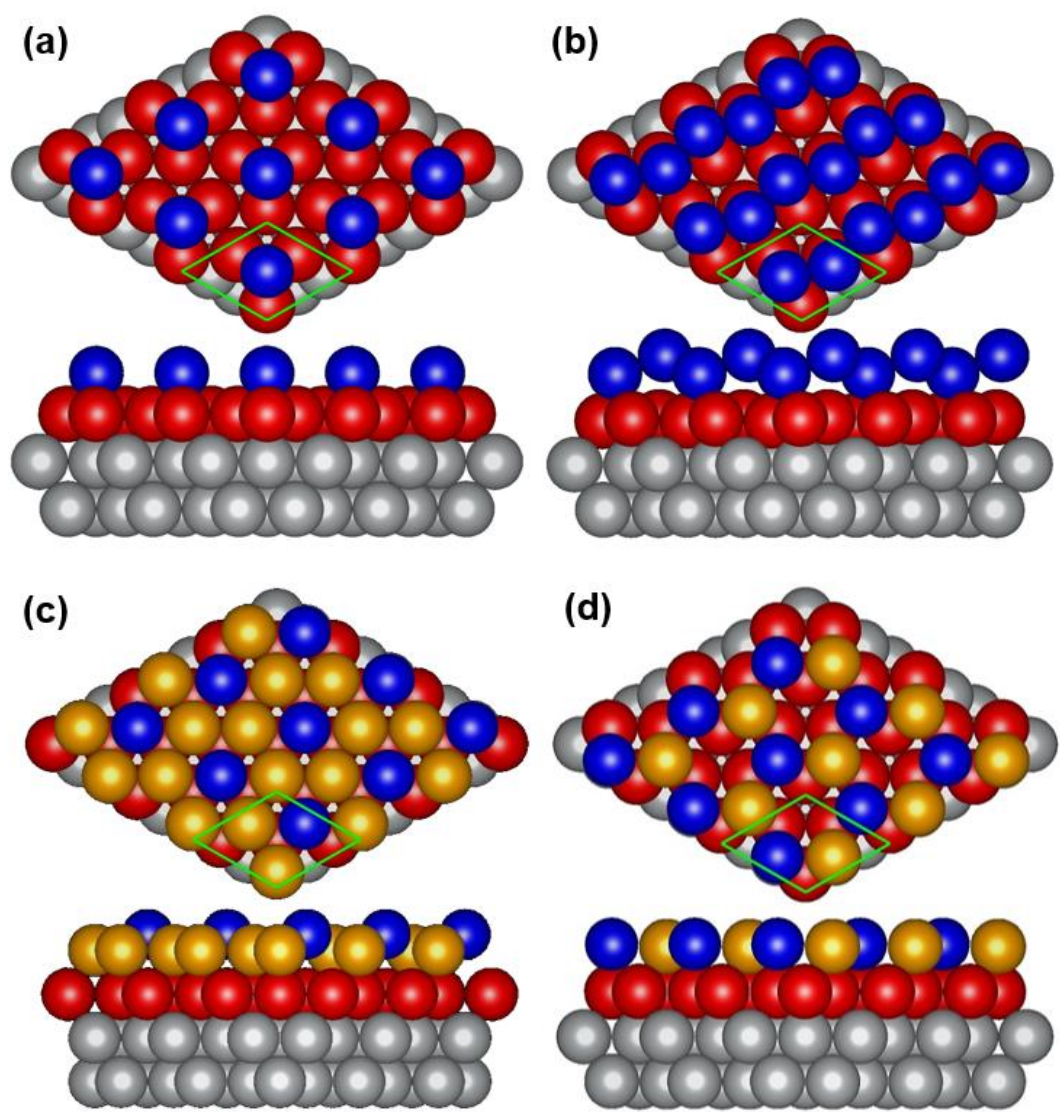

**Figure S1 Top and side views of the investigated atomic models of Te/Ag(111)√3×√3 after full relaxation.** (a) One Te atom (blue) per √3×√3 cell (green lines) is positioned at a hollow site of the Ag(111) surface (red). (b) A monolayer of tellurene was initially placed on the Ag(111) surface as an adlayer. The honeycomb structure of tellurene was distorted and a buckling of 0.95 Å resulted from the relaxation as illustrated in the figure. (c) One out of three Ag atoms of the upper layer (yellow) is substituted by one Te atom in a periodic fashion to result in a √3×√3 unit cell. After relaxation, the Te atoms were located 0.87 Å higher than the Ag atoms. (d) The binary Te-Ag honeycomb alloy discussed in the manuscript. The Te-Ag alloy stayed flat during relaxation. Nine layers of Ag atoms were used in the calculations, three of them are shown in the figures (red+gray). All models are shown after full relaxation using the VASP package.

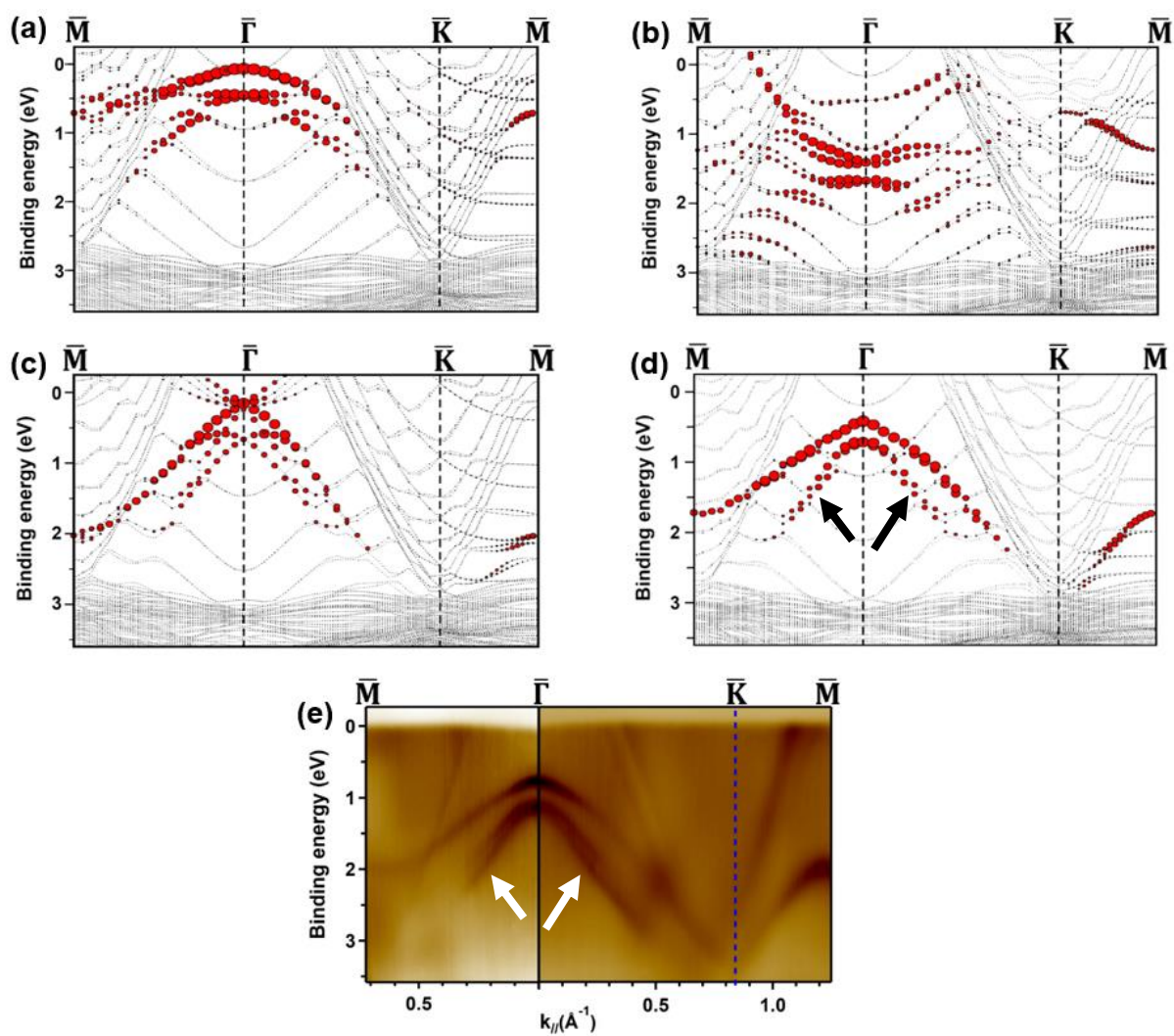

**Figure S2 Band structure calculations of the models in figure S1 and ARPES data.** Panels (a)-(d) show the theoretical band structures obtained for models (a)-(d) in Fig. S1, respectively. (e) Experimental band structure data plotted along the same symmetry lines as for the theoretical band structures for an easy comparison. One can clearly see that the calculated band structure of the binary Te-Ag honeycomb alloy, model (d), fits the experimental data very well, both regarding the shapes of the dispersions and the band widths. Also the absolute positions of the bands are in good agreement with the experimental results. Furthermore, the small split of the S2 band is reproduced by the theoretical band structure, see arrows in (d) and (e).